\documentclass[11pt,a4paper]{article} 
\usepackage{jheppub}

\usepackage{latexsym} 	
\usepackage{amssymb}  	
\usepackage{amsbsy}
\usepackage{amsmath}

\numberwithin{equation}{section}

\newcommand{\Tr}{\mbox{Tr}\,}

\newcommand{\bra}{\langle}
\newcommand{\ket}{\rangle}
 
\newcommand{\half}{\frac{1}{2}}
\newcommand{\re}{\mbox{Re\,}}
\newcommand{\im}{\mbox{Im\,}}
\newcommand{\eps}{\epsilon}

\newcommand{\be}{\begin{equation}}
\newcommand{\ee}{\end{equation}}
\newcommand{\bea}{\begin{eqnarray}}
\newcommand{\eea}{\end{eqnarray}}
\newcommand{\bean}{\begin{eqnarray*}}
\newcommand{\eean}{\end{eqnarray*}}

\title{Some remarks on Lefschetz thimbles and complex Langevin dynamics}

\author[a]{Gert Aarts,}
\author[a]{Lorenzo Bongiovanni,}
\author[b]{Erhard Seiler}
\author[c]{and D\'enes Sexty}

\affiliation[a]{Department of Physics, College of Science, Swansea University, Swansea, United Kingdom}
\affiliation[b]{Max-Planck-Institut f\"ur Physik (Werner-Heisenberg-Institut), M\"unchen, Germany}
\affiliation[c]{Institut f\"ur Theoretische Physik, Universit\"at Heidelberg, Heidelberg, Germany}

\emailAdd{g.aarts@swan.ac.uk}
\emailAdd{pylb@swan.ac.uk} 
\emailAdd{ehs@mppmu.mpg.de}
\emailAdd{d.sexty@thphys.uni-heidelberg.de}

\abstract{
 Lefschetz thimbles and complex Langevin dynamics both provide a means to tackle the numerical sign problem prevalent in theories with a complex weight in the partition function, e.g.\ due to nonzero chemical potential. Here we collect some findings for the quartic model, and for U(1) and SU(2) models in the presence of a determinant, which have some features not discussed before, due to a singular drift. We find evidence for a relation between classical runaways and stable thimbles, and give an example of a degenerate fixed point. We typically find that the distributions sampled in complex Langevin dynamics are related to the thimble(s), but with some important caveats, for instance due to the presence of unstable fixed points in the Langevin dynamics.
 }


\arxivnumber{1407.2090}

\begin{document}
\maketitle

%%%%%%%%%%%%%%%%%%%%%%%%%%%%%%%%%%%%%%%%%%%%%%%%%%%

\section{Introduction}
\label{sec:intro}

Recently there has been stimulating progress in attempts to  solve the numerical sign problem in QCD and related theories at nonzero baryon density, by allowing the field variables to take value in the complexification of the original configuration space, see e.g.\ Refs.\ 
\cite{Aarts:2013bla,Aarts:2013uxa,Cristoforetti:2013qaa} for reviews. The idea here is obvious: while the sign problem may be severe in the original formulation, due to a complex and highly oscillatory integrand in the path integral,  an extension into the complexified field space may  ameliorate or even eliminate the sign problem altogether, a possibility well known from examples of simple integrals with complex saddle points.

The main approaches which have been pursued are complex Langevin dynamics 
\cite{Parisi:1984cs,Klauder:1983,Damgaard:1987rr,Aarts:2008rr,Aarts:2008wh,arXiv:0912.3360,Aarts:2010gr,arXiv:1101.3270,Aarts:2009dg,Aarts:2012ft,Seiler:2012wz,Sexty:2013ica,Bongiovanni:2013nxa,Duncan:2012tc,Aarts:2013uza,Mollgaard:2013qra,Fromm:2011qi,Fromm:2012eb,Langelage:2014vpa,Greensite:2014cxa}
and integration along Lefschetz thimbles 
\cite{Witten:2010cx,Cristoforetti:2012su,Cristoforetti:2013wha,Mukherjee:2013aga,Fujii:2013sra,Cristoforetti:2014gsa,Mukherjee:2014hsa,Tanizaki:2014xba,Aarts:2013fpa}. 
In the latter, the original path of integration is deformed in order to pass through the fixed (or critical) points of the complex action, which typically reside in the complexified space. The contour then follows paths of steepest descent, along which the imaginary part of the action is constant, the so-called thimbles. 
 In this approach the sign problem is not eliminated, but is in fact replaced by two sign problems with a different origin. First of all, there is a so-called residual sign problem along each thimble, due to the curvature of the integration contour in the complexified space, i.e.\ the complex Jacobian present when an explicit parametrisation of the thimble is introduced. A second sign problem appears in the case that more than one thimble contributes, and the thimbles, with differing imaginary parts of the action associated with the different critical points, have to be combined including their respective phases. This is typically referred to as a global sign problem. 
The numerical implementation of this approach \cite{Cristoforetti:2012su} has been tested in a number of models, including the interacting Bose gas at nonzero chemical potential in four dimensions
\cite{Cristoforetti:2013wha,Fujii:2013sra}, 
which has a severe sign problem \cite{Aarts:2008wh}, and current advances have focussed on better control over the residual sign factor in the case of a single thimble \cite{Fujii:2013sra,Cristoforetti:2014gsa}.\footnote{For related studies on the analytic continuation of path integrals, see e.g. Refs.\ \cite{Guralnik:2007rx,Ferrante:2013hg,Basar:2013eka,Cherman:2014ofa}.}

In  complex Langevin dynamics, or stochastic quantisation, the complexified field space is explored stochastically, such that a real and nonnegative distribution is effectively sampled during the Langevin process. While this method was proposed some time ago \cite{Parisi:1984cs,Klauder:1983}, only recently it has been shown convincingly that it can handle severe sign problems \cite{Aarts:2008wh,Aarts:2010gr} and that numerical problems of the past, such as runaways, can be eliminated \cite{Aarts:2009dg}. Moreover, the theoretical foundation has been clarified \cite{arXiv:0912.3360,arXiv:1101.3270}. The control over nonabelian gauge theories has been drastically improved with the implementation of gauge cooling \cite{Seiler:2012wz}, possibly adaptively \cite{Aarts:2013uxa},  and first results for QCD at nonzero baryon chemical potential have appeared \cite{Sexty:2013ica}. Another recent application is to SU(3) gauge theory in the presence of a nonzero theta-term \cite{Bongiovanni:2013nxa}.

Both methods are not without open questions, though. In the thimble approach, both the residual and the global sign problems require further study, as does the implementation for gauge theories. In complex Langevin dynamics, the applicability of the method is determined by the localisation of the distribution in the complexified space  \cite{arXiv:0912.3360,arXiv:1101.3270} (see e.g.\ Ref.\ \cite{Aarts:2013uza} for a recent explicit study)
and by the holomorphicity of the original action, which is in principle violated in the presence of a determinant in the Boltzmann weight and might lead to problems \cite{Mollgaard:2013qra} (see also Ref.\ \cite{Greensite:2014cxa}).

Since both methods explore the complexified field space, it is worth investigating the relation between the two. First steps in this direction were taken in Ref.\ \cite{Aarts:2013fpa}. Here we extend that analysis by studying additional models with features ultimately relevant for QCD, namely U(1) and SU(2) models in the presence of a determinant. We find that those are also of interest for the thimble approach, since thimbles can end at the zeroes of the determinant, a situation not encountered before. 
 This paper is organised as follows. In the following section we make some general remarks, using the quartic model as an example.
The U(1) model, previously discussed in Refs.\ \cite{Aarts:2008rr,Mollgaard:2013qra}, 
is considered in Sec.\ \ref{sec:onelink},
while the SU(2) model of Refs.\ \cite{Aarts:2013uxa,Aarts:2012ft,Berges:2007nr} is further analysed in Sec.\ \ref{sec:su2}, also in combination with gauge cooling. The final section concludes.

\section{Thimbles and Langevin dynamics}
\label{sec2}

In Ref.\  \cite{Aarts:2013fpa} a first comparison between Lefschetz thimbles and complex Langevin dynamics was made for the quartic model in zero dimensions. Here we provide some more general remarks. We consider a holomorphic complex action $S(z)$, with the partition function defined by integration along the real axis,\footnote{For sake of simplicity, we use the notation of one degree of freedom, but all equations can be readily extended to higher dimensions.}
\be
Z = \int_{-\infty}^\infty dx\, e^{-S(x)}.
\ee
Since the action is complex, there is a numerical sign problem.
 For illustration purposes, we use again the quartic model, but now with a linear term, to explicitly break the symmety $z\to -z$, 
\be
\label{eqact}
S(z) = \frac{\sigma}{2} z^2+\frac{1}{4}z^4+hz,  \quad\quad\quad h\in \mathbb{C}.
\ee
The complexity is in this case  introduced by the linear term, while the coefficient of the quadratic term $\sigma$ is taken as unity (for a detailed study with $h=0$ but a complex $\sigma$, see Ref.\ \cite{Aarts:2013uza}).

In complex Langevin dynamics, the sign problem is tackled by solving the Langevin equation in the complex plane, 
\be
\dot z = -\partial_z S(z) +\eta,
\ee 
or,  writing $z=x+iy$,
\be
\dot x  = - \re\partial_z S(z) +\eta,
\quad\quad\quad
\dot y =  -\im\partial_z  S(z).
\ee
Here $\eta$ is (real) noise with variance 2, $\bra\eta^2\ket=2$. 
 The partition function is then represented as
\be
Z = \int dxdy\, P(x,y),
\ee
where $P(x,y)$ is formally given as the solution of the associated Fokker-Planck equation and in practice constructed during the Langevin evolution.
The justification of this approach, together with criteria for correctness, is discussed in detail in Refs.\  \cite{arXiv:0912.3360,arXiv:1101.3270}.
We refer to the equations without noise as the classical Langevin equations.

Thimbles, on the other hand,  are given by a deformation of the original integration contour, determined by first finding the critical or fixed points $z_k$, where $\partial_z S=0$, and then constructing the submanifolds passing through a fixed point for which $\im S(z)=$ constant.
Using the Cauchy-Riemann equations and parameterising the thimble with a parameter $t$, leads to an evolution equation for the (stable) thimble ${\cal J}_k$, the curve of steepest descent, 
\be
\dot z = -\overline{\partial_z S(z)},
\ee
or
\be
\label{eq27}
\dot x  = - \re\partial_z  S(z),
\quad\quad\quad
\dot y =  +\im\partial_z  S(z).
\ee
where $z\to z_k$ as $t\to\infty$.
Unstable thimbles ${\cal K}_k$, the curves of steepest ascent,  are obtained by reversing the sign of $t$.  
In this case the partition function is written as a sum over the stable thimbles,
\be
 Z = \sum_k m_k e^{-i{\rm Im} S(z_k)}\int_{{\cal J}_k} dz\, e^{-{\rm Re} S(z)}.
 \ee
 Here $m_k$ are the intersection numbers (see below).
 Integrating along these stable thimbles leaves both a residual sign problem, due to the curvature of the thimble, and a global sign problem, in the case that more than one thimble contributes. 

 The thimble equations are the complex conjugate of the classical Langevin equations, with an opposite sign for the drift in the imaginary direction. This has various consequences. First of all, in the case of classical Langevin dynamics, fixed points are either attractive or repulsive (we always assume the nondegenerate case, $\partial_z^2S|_{z=z_k}\neq 0$, except in Sec.\ \ref{sec43}). Instead, under the thimble evolution  fixed points have a stable and unstable direction, which can be seen by linearising around $z_k=x_k+iy_k$ to find
\be
\left(
\begin{array}{c} \dot x \\ \dot y \end{array}
\right)
= -
\left(
\begin{array}{rr} 
h_{k1} & h_{k2} \\
h_{k2} & -h_{k1} 
 \end{array}
\right)
\left(
\begin{array}{c} x -x_k\\ y-y_k \end{array}
\right),
\ee
with
\be
h_{k1}= \partial_x^2\re S\big|_{z=z_k}, \quad\quad\quad\quad h_{k2} = \partial_x\partial_y \re S\big|_{z=z_k},
\ee
where the Cauchy-Riemann equations have been used. The eigenvalues of the Hessian are indeed real and opposite. The eigenvectors in the complex plane correspond to the directions of the stable and the unstable thimble.

\begin{figure}[t]
\begin{center}
\epsfig{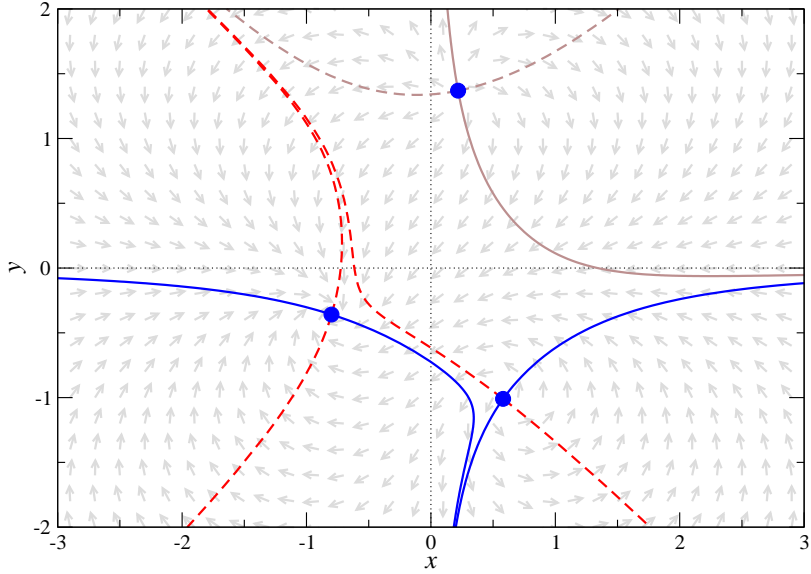}
\end{center}
\caption{Thimbles and Langevin flow in the quartic model with $\sigma=1$ and $h=1+i$: the blue circles denote the fixed points, the (normalised) arrows the classical Langevin drift, and full (dashed) lines the stable (unstable) thimbles. The two blue thimbles contribute. The third fixed point does not contribute.
}
 \label{fig:quartic1}
\end{figure}

This is illustrated in Fig.\  \ref{fig:quartic1} for the quartic model (\ref{eqact}), where the fixed points are indicated with blue circles and the classical Langevin drift with (normalised) arrows, for a specific choice of parameters, $\sigma=1$, $h=1+i$. The fixed points are determined by 
\be
z^3+\sigma z +h=0, 
\ee
and given by the three roots
\be
z_k = e^{2\pi i k/3}D - e^{-2\pi ik/3} \frac{\sigma}{3D},  \quad\quad\quad k=0,1,2,
\ee
where
\be
D = \left( -\frac{h}{2}+\frac{h}{2}\sqrt{1 + \frac{4\sigma^3}{27h^2}}\right)^{1/3}.
\ee 
For $\sigma=1$ and $h=1+i$ this gives, under Langevin evolution, the stable fixed point
$z_0 = -0.799 -i 0.359$ and the unstable fixed points $z_1=
0.219+i1.369$ and $z_2=0.580 -i 1.010$.

Every fixed point has an associated stable thimble, indicated with a full line, and an unstable thimble, indicated with a dashed line.\footnote{We construct these thimbles using numerical integration of Eq.\ (\ref{eq27}), starting very close to the fixed point.} 
We note that the stable thimbles end asymptotically in the region where the original integral converges (i.e.\ where $z^4>0$), while the unstable thimbles end in the region of nonconvergence ($z^4<0$).
The oriented intersection numbers $m_k$ are defined from the crossing of the unstable thimble with the original domain of integration. 
These numbers make sure that the integration along the thimbles connects the convergence region on the negative real axis to the convergence region on the positive real axis (with possibly visiting other convergence regions first).
 Hence for the two lower thimbles, $m_{0,2}=1$, while for the upper thimble $m_1=0$. The contributing thimbles are then given by the blue full lines and are seen to be a deformation of the original contour of 
   integration.\footnote{We note here that the fact that two thimbles contribute is a consequence of the Stokes phenomenon \cite{Berry}. When $h$ is real, there is one thimble (the real axis) and one contributing fixed point, which is on the real axis. Making $h=|h| e^{i\theta}$ slightly complex by increasing $\theta$ from $0$, deforms the contour into the complex plane. For certain values of $h$ the deformation will be such that the first thimble passes through a second critical point as well. The values of $h$ where this occurs form the Stokes lines in the complex $h$ plane and are determined by $\im S(z_i) = \im S(z_j)$. For larger $\theta$ two thimble contribute, as in Fig.\ \ref{fig:quartic1}. For $|h|=\sqrt{2}$, the angle on the Stokes line is $\theta=0.7323$, just below $\pi/4=0.7854$, which is  used in Fig.\ \ref{fig:quartic1}.}
   We have verified that integration along the two thimbles yields the correct results, provided that the complex Jacobian $z'(t)$ is included.\footnote{The partition function is given by the sum of the two thimble contributions associated with $z_0$ and $z_2$, which are $Z_0 = 1.744 + i0.461$ and $Z_2 = 0.021 + i0.426$. The sum, $Z=Z_0+Z_2 = 1.765 + i0.887$, is the correct answer. We note that for the real part the first thimble is dominant, while for the imaginary part both thimbles contribute equally.}
 We note that in this case there is both a global sign problem, since $\im S(z_0)\neq \im S(z_2$), as well as a residual sign problem for each thimble.

The first observation is therefore that both an attractive and a repulsive fixed point under Langevin dynamics contribute to the thimble contour. We come back to this below.
A comparison between the classical Langevin drift and the stable thimbles reveals another relation.  In Langevin dynamics one frequently encounters runaways trajectories, along which the classical evolution diverges to infinity in a finite time. These trajectories are associated with a repulsive fixed point.  In this case this occurs at $x=0$, where for large $|y|$ the classical Langevin equation simplifies to $\dot y = y^3$. In the stochastic evolution, these instabilities are regulated by the noise, although in some cases a careful adaptive numerical integration is required \cite{Aarts:2009dg}. Since the $y$ component of the drift in the thimble evolution has the opposite sign,
we observe that  for large $|y|$  the stable thimble in fact coincides with a runaway trajectory in the Langevin evolution.\footnote{Here it should be noted that the weight on the thimble for large $|y|$ is exponentially small and hence this region hardly contributes. Moreover, the contributions from both thimbles effectively cancel, due to the opposite orientation.}
This should be contrasted with the situation at large $|x|$ and $y\sim 0$, where the flow along the stable thimble and the Langevin drift point in the same direction, namely towards the stable fixed point.

\begin{figure}[t]
\begin{center}
\epsfig{figure=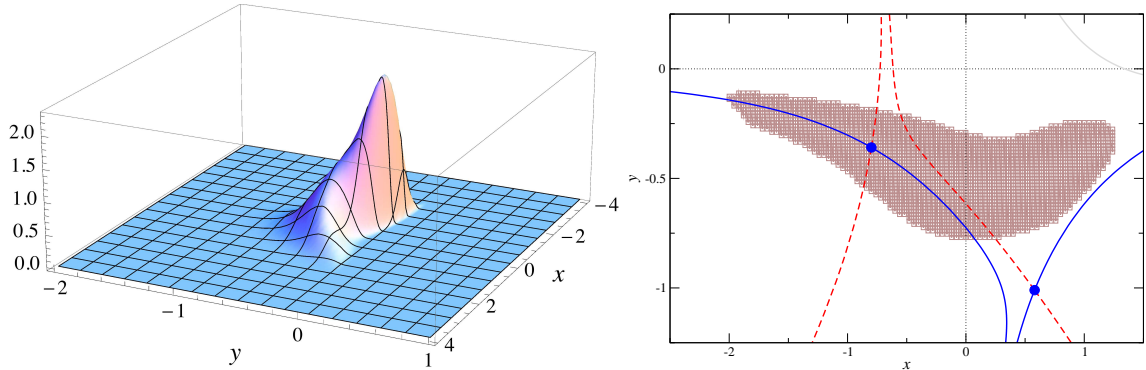,width=0.53\textwidth}
\epsfig{figure=plot-flow-th-hx-hist.eps,width=0.45\textwidth}
\end{center}
\caption{Quartic model, with $\sigma=1$ and $h=1+i$. Histogram collected during a complex Langevin simulation (left) and a comparison with the thimbles (right). 
}
 \label{fig:quartic2}
\end{figure}

In order to compare the Langevin distribution with the thimbles, we have collected the histogram $P(x,y)$ during a Langevin simulation. The result for $\sigma=1$ and $h=1+i$ is given in Fig.\  \ref{fig:quartic2} (left). 
Although the distribution appears to be localised, there is a power decay at large distances, which leads to incorrect convergence for higher moments $\bra z^n\ket$. This situation has been discussed in great detail in Ref.\ \cite{Aarts:2013uza} for the model with $h=0$ and complex $\sigma$, and in fact that analysis carries over immediately to the case discussed here, 
with complex $h$. Hence we will not discuss it further. 
A comparison between the Langevin distribution and the thimbles is shown in  Fig.\  \ref{fig:quartic2} (right). We note that the Langevin distribution is stretching along the thimble associated with the attractive fixed point under Langevin evolution. 
However, since the second thimble is associated with a repulsive fixed point under Langevin evolution, it is avoided, making the regions explored in the complex plane manifestly distinct.
We conclude therefore that there is less overlap between the Langevin distribution and the contributing thimbles than previously 
found in Ref.\ \cite{Aarts:2013fpa}, where only a single thimble contributed.

\section{U(1) model with a determinant}
\label{sec:onelink}

We now extend the study of thimbles to models with a determinant in the weight and hence the logarithm of the determinant in the effective action. Both in Langevin dynamics and thimble dynamics this leads to a qualitatively new feature, namely a singular drift where the determinant vanishes in the complexified configuration space. Formally this will lead to a breakdown of the justification of Langevin dynamics \cite{arXiv:0912.3360,arXiv:1101.3270}, due to the lack of holomorphicity, and it can lead to a wrong convergence in practice \cite{Mollgaard:2013qra}, whereas in thimble dynamics this situation has not been studied before, as far as we know.

We start with a U(1) model with one link, written as $U=e^{ix}$. The partition function we consider is motivated by QCD at nonzero chemical potential and was introduced in Ref.\ \cite{Aarts:2008rr}. It takes the form

 \be
 Z = \int_{\rm U(1)} dU \, e^{-S_B}  \det M
 = \int_{-\pi}^{\pi}\frac{dx}{2\pi}\, e^{\beta\cos x} \left[ 1+\kappa \cos(x-i\mu)\right].
\ee
We take $\beta$ and $\kappa$ real and positive, such that the complex weight is introduced via the chemical potential $\mu$. 
The determinant satisfies $[\det M(\mu)]^* = \det M(-\mu^*)$. When $\kappa<1$, the weight is real and positive when $\mu=0$, while for $\kappa>1$ there is already a real sign problem at $\mu=0$.
For $\kappa<1$ this model was studied in detail in Ref.\ \cite{Aarts:2008rr}, while for $\kappa>1$ possible issues with the determinant were noted in Ref.\ \cite{Mollgaard:2013qra}. An analytical evaluation yields
\be
Z = I_0(\beta) + \kappa I_1(\beta)\cosh\mu,
\ee
where $I_n(\beta)$ are the modified Bessel functions of the first kind.

We hence consider the (effective) action
\be
 S(z)  = -\beta\cos z - \ln \left[  1+\kappa\cos(z-i\mu)\right],
\ee
where the principle branch of the logarithm is understood,
with the drift
\be
-\partial_z S(z)  = -\beta\sin z - \frac{\kappa\sin(z-i\mu) }{ 1+\kappa\cos(z-i\mu)}.
\ee
Fixed points are determined by $\partial_z S(z)=0$ and singular points by $1+\kappa\cos(z-i\mu)=0$. We have to make a distinction between $\kappa$ greater or less than 1. Consider first $\kappa<1$. Then it is easy to see that under Langevin flow there is one attractive fixed point at $x=0$, while the repulsive fixed points are at $x=\pm\pi$. The drift is singular at $z=\pm\pi+iy_s$, with $\cosh(y_s-\mu)=1/\kappa$.
On the other hand,  when $\kappa>1$, some of the repulsive fixed points move away from $x=\pm\pi$, while also the singular drift is no longer at $x=\pm\pi$, but instead at $z=x_s+i\mu$, with $\cos x_s = -1/\kappa$.

Similarly, some of the thimbles can be found analytically. Consider the complex weight
\be
w(z) = e^{-S(z)} = \left[1+\kappa \cos(z-i\mu)\right] e^{\beta\cos z}.
\ee
For $x=0$
the corresponding action is real for all $y$, hence the $x=0$ axis corresponds to the unstable thimble associated with the fixed point on the $x=0$ axis (the instability follows from the Hessian, but also from the fact  that $z\to\pm i\infty$ is not in the region of convergence of the original integral).
At $x=\pm\pi$ the weight equals
\be
w(\pm\pi+iy)  = \left[1-\kappa \cosh(y-\mu)\right] e^{-\beta\cosh y},
\ee
and hence the nature of the thimble depends again on $\kappa$. 
For $\kappa>1$ the determinant is always negative and the action has a constant imaginary part,  $\im S=\pm\pi$, for all $y$.
When $\kappa<1$, the thimble's properties depend on $\mu$.  When $\kappa \cosh(y-\mu) < 1$, the determinant is real and positive and hence $\im S=0$. On the other hand,
when $\kappa \cosh(y-\mu) > 1$,  the determinant is real and negative and $\im S=\pm\pi$. These two cases are separated by the singularity in the drift at $\kappa \cosh(y-\mu)=1$. Hence the imaginary part of the action changes by a constant at this singularity.
Finally, there are also thimbles not at $x=0,\pm\pi$, but these cannot be given in analytical form.
However, here it should be noted that around the singular points, the imaginary part of the action varies by $2\pi$, and hence it is expected that thimbles may end at those singular points.

\begin{figure}[t]
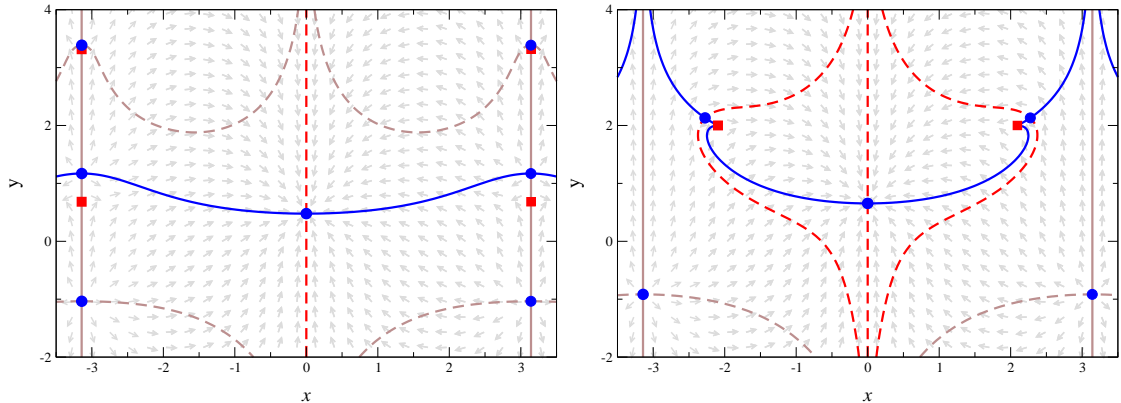

\begin{center}
\epsfig{figure=plot_full_b1_k0.25_m2-th-v3.eps,width=0.48\textwidth}
\epsfig{figure=plot_full_b1_k2_m2-th-v3.eps,width=0.48\textwidth}
\end{center}
 \caption{Thimbles and Langevin flow in the U(1) one-link model  with $\beta=1,\mu=2$ and $\kappa=1/2$ (left) and $\kappa=2$ (right): the blue circles indicate the fixed points,  the (normalised) arrows the classical Langevin drift, and full (dashed) lines the stable (unstable) thimbles. Note that the blue line on the left is the stable (unstable) thimble for the fixed point at $x=0$ ($x=\pm\pi$).   The squares indicate where the flow diverges, $\im S$ jumps, and the direction of the flow along the thimble changes sign. Only the blue thimble(s) contribute. 
}
 \label{fig:flow_full_4}
\end{figure}

This is best illustrated in a plot, see Fig.\ \ref{fig:flow_full_4}. For definiteness, we take $\beta=1$, $\mu=2$ and $\kappa=1/2$ (left) and 2 (right).
Starting from the fixed point at $x=0$, the vertical red line is the unstable thimble and the blue line is the stable thimble. The stability of the other thimbles can be understood by following the flow along the thimbles (or from the Hessian of course). Note that for $\kappa=1/2$, the blue line is the stable thimble for the fixed point at $x=0$, but the unstable thimble for the fixed points at $x=\pm\pi$. The stable thimbles at $x=\pm\pi$, going to $y=\pm\infty$, correspond again to runaway trajectories under Langevin dynamics.
We note that the original integral converges when $z\to \pm\pi \pm i\infty$, but not when  $z\to \pm i\infty$.
At the red squares, the drift diverges, the imaginary part of the action jumps and the direction of the flow along the thimble changes sign. Note that there are two singular points at $x=\pm\pi$ for $\kappa<1$.
Those singular points  merge at $\kappa=1$ and then move away from $x=\pm\pi$ as $\kappa>1$. In this process the central fixed point at $x=\pm\pi$ is absorbed and only one fixed point at $x=\pm\pi$ remains. Instead of one contributing thimble, there are now two additional contributing thimbles (related by symmetry) going out to $z=\pm\pi+ i\infty$, as in the example in the preceding Section. 
For $\kappa>1$, the imaginary part of the action jumps at the singular point, but not by $\pi$ as is the case when $\kappa<1$. This can also be seen by observing that the thimbles do not reach
the singular point in opposite directions.\footnote{For the parameter values used in Fig.\ \ref{fig:flow_full_4} (right), we find that at the fixed point $z_1=i0.654$, $\im S= 0$, while at the other fixed points ($z_{2,3}=\pm 2.28 + i2.13$), $\im S= \mp 2.59$.}

\begin{figure}[t]
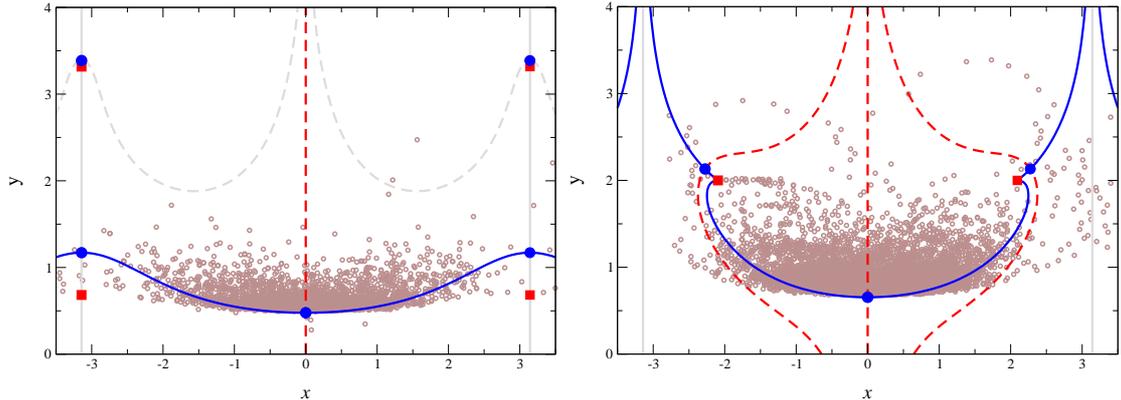

\begin{center}
\epsfig{figure=plot_full_b1_k0.25_m2-th-v4.eps,width=0.48\textwidth}
\epsfig{figure=plot_full_b1_k2_m2-th-v4.eps,width=0.48\textwidth}
\end{center}
 \caption{As in the previous plot, with scatter data from a complex Langevin simulation added.
}
 \label{fig:flow_full_5}
\end{figure}

We denote the critical points as $z_k$ and the corresponding stable thimbles as ${\cal J}_k$. For each contributing thimble we then have the partition function
\be
 Z_k = \int_{{\cal J}_k} dz\, e^{-S(z)} = e^{-i{\rm Im} S(z_k)} \int_{{\cal J}_k} dz\, e^{-{\rm Re} S(z)},
\ee
and the full partition functions is the sum,
\be
Z = \sum_k m_kZ_k,
\ee
with $m_k$ the intersection number. In the case of Fig.\ \ref{fig:flow_full_4} (left) there is only one contributing thimble; we have verified that integrating over this thimble yields the correct result, provided that the residual sign problem is taken into account. 
In the case of Fig.\ \ref{fig:flow_full_4} (right), the three fixed points $z_{1,2,3}$ all contribute (with intersection number 1) and 
we find, including the global and residual phases,  the following contributions to the partition function,
\be
Z_1=	34.6686, \quad\quad Z_2=Z_3^*=0.0027 - i0.0115.
\ee
The complete partition function is given by the sum, $Z=Z_1+Z_2+Z_3=34.6740$, which is indeed the correct answer.
We observe that the first thimble is dominant.

It is now of interest to compare the thimble structure with the results from a complex Langevin simulation. In Ref.\ \cite{Aarts:2008rr}, the dynamics under Langevin evolution was studied with scatter plots (for $\kappa=1/2<1$). Here we take the data from that reference and compare it with the thimbles in Fig.\ \ref{fig:flow_full_5} (left). We note that the Langevin data stays close to the thimble, with the occasional excursion to larger $y$ values. Since there is noise in the $x$ direction (real noise), the distribution is spread horizontally and hence does not follow the curvature of the thimble. Expectation values for the lowest moments $\bra e^{inz}\ket$, with small $n$, are correctly reproduced \cite{Aarts:2008rr}.
 We have generated new Langevin data for $\kappa=2$, see Fig.\ \ref{fig:flow_full_5} (right). Here we observe a similar pattern as at $\kappa<1$, however with much larger excursions in the $y$ direction. In this case, the correct results for the lowest negative moments are not reproduced. It is tempting to associate the failure of Langevin in this case with the appearance of the thimbles stretching out $z=\pm\pi+ i\infty$, as well as with the presence of the determinant. However, at this stage a complete understanding along the lines of  Refs.\   \cite{arXiv:0912.3360,arXiv:1101.3270} is still lacking and hence we defer further discussion to a future publication. 

To summarise, we find that in the presence of a determinant the flow has singular points. For Langevin dynamics, this leads to a breakdown of the formal justification and possible wrong results in practice. For the Lefschetz approach, we find that thimbles may end at singular points and the imaginary part of the action jumps by a constant.
Hence, if there is more than one contributing thimble, they connect either at $|z|\to\infty$ or at a singularity.

\section{SU(2) model}
\label{sec:su2}

We now extend the analysis to the simple nonabelian one-link model,
\be
Z = \int_{\rm SU(2)} dU \exp\left[\frac{\beta}{2}\Tr U\right],
\ee
with complex $\beta$ and a gauge symmetry, $U\to \Omega U\Omega^{-1}$, where $U, \Omega\in$ SU(2). The exact result is $Z=  I_1(\beta)/\beta$. For Langevin dynamics we can follow (at least) two approaches \cite{Aarts:2012ft}: a complete `gauge fixing', which will lead to a logarithm in the effective action due to the reduced Haar measure, and matrix updates combined with gauge cooling \cite{Seiler:2012wz}, which can readily be extended to full nonabelian gauge theories in four dimensions with dynamical fermions \cite{Sexty:2013ica}. It is therefore interesting to compare both approaches with the Lefschetz thimbles.

\subsection{Complete gauge fixing}

In this approach the partition function is written as
\be
Z = \int_{-\pi}^\pi \frac{dx}{2\pi}\, \sin^2x\, e^{\beta\cos x} =  \int_{-\pi}^\pi \frac{dx}{2\pi}\, e^{-S(x)} , 
\ee
with
\be
S(z) = -\beta\cos z -2\ln\sin z.
\ee
We note that the reduced Haar measure leads to a logarithm in the action. 
The drift is given by  
\be
-\partial_z S(z)  = -\beta \sin z +2\frac{\cos z}{\sin z}, 
\ee
and hence the fixed points are at
\be
\cos z_\pm = -\frac{1}{\beta}\left(1\pm\sqrt{1+\beta^2}\right).
\ee
Due to the logarithm, the drift has singular points at  $\cos z = \pm 1$. 
The second derivative at the fixed points equals
\be
\partial_z^2 S(z)\big|_{z=z_\pm}  = \pm 2\sqrt{1+\beta^2},
\ee
hence there is a degeneracy at $\beta^2=-1$, which we discuss below.
Finally, the action at the fixed points is given by
\be
S(z_\pm) = 1\pm \sqrt{1+\beta^2}-\ln\left(1\pm\sqrt{1+\beta^2}\right) +\ln(-\beta^2/2).
\ee

\begin{figure}[t]
\begin{center}
\epsfig{figure=plot-SU2-thimble-xy.eps,width=0.48\textwidth}
\epsfig{figure=plot-SU2-thimble-action.eps,width=0.48\textwidth}
\end{center}
\caption{
SU(2) one-link model, with $\beta=(1+i\sqrt{3})/2$: thimbles in the $xy$ plane (left) and the $\Tr U$ plane (right). 
Blue circles indicate fixed points, full (dashed) lines the stable (unstable) thimbles, and red squares the singular points. The blue thimble contributes.
}
 \label{fig:SU2-th}
%\end{figure}
%\begin{figure}[t]
\begin{center}
\epsfig{figure=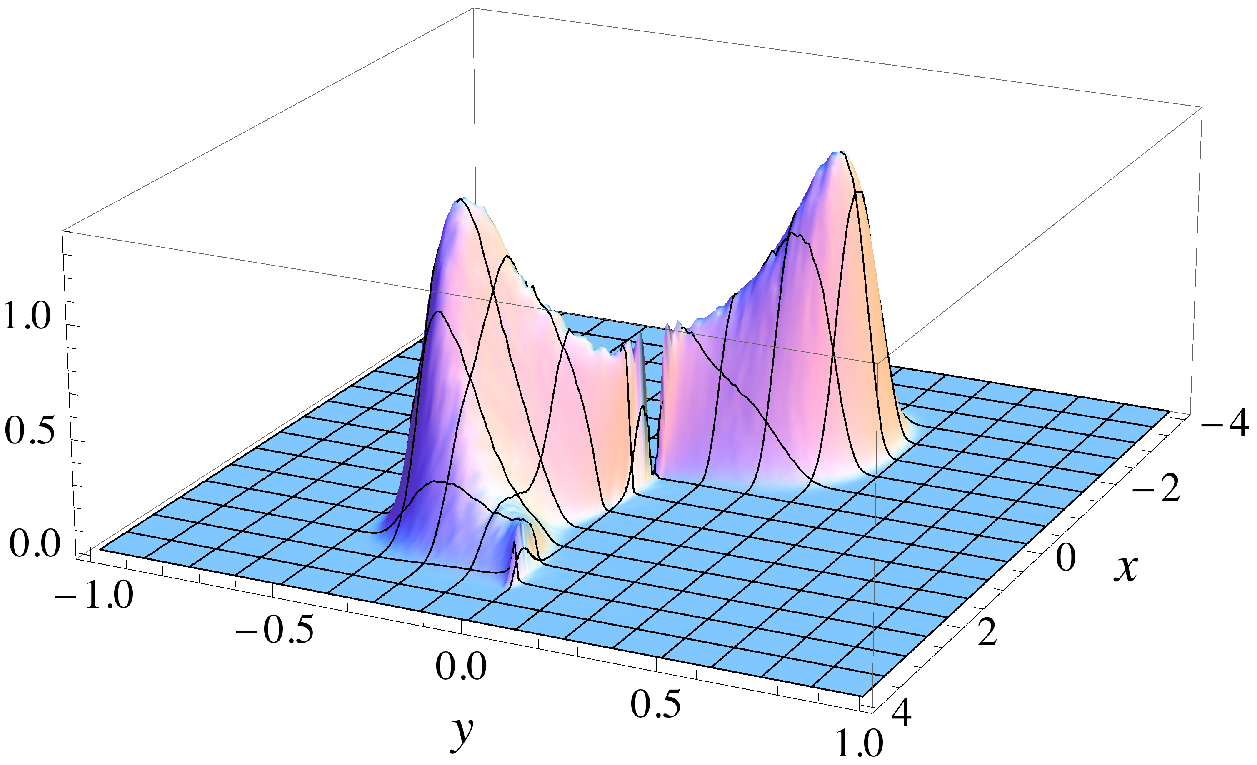,width=0.48\textwidth}
\epsfig{figure=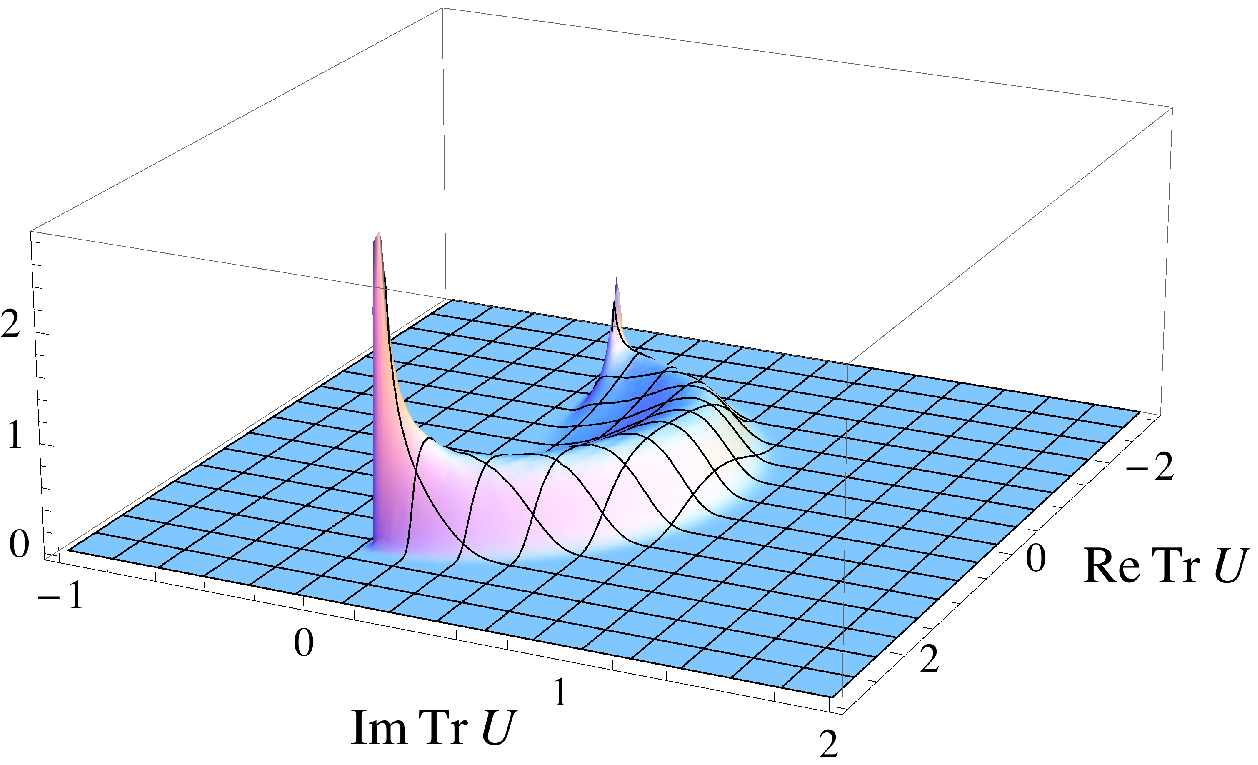,width=0.48\textwidth}
\end{center}
\caption{
Histograms collected during a complex Langevin simulation in the gauge fixed formulation of the  SU(2) one-link model, with $\beta=(1+i\sqrt{3})/2$, in the $xy$ plane (left) and the $\Tr U$ plane (right).
}
 \label{fig:SU2-CL}
\end{figure}

The thimbles can be found by numerical integration, as above. 
In Fig.\ \ref{fig:SU2-th} (left) the thimbles are shown for $\beta=(1+i\sqrt{3})/2$. As above, the blue circles (red squares) denote the fixed (singular) points. The full blue line is the contributing thimble. Note the symmetry $z\to-z$. The thimble through the upper (and lower) fixed point does not contribute, since the intersection number vanishes. As in the U(1) case,  thimbles end at the singular points in the drift. 
In order to compare with the gauge invariant approach below, we also show the thimbles in the $\Tr U = 2\cos z$ plane, see 
Fig.\ \ref{fig:SU2-th} (right).

\begin{figure}[t]
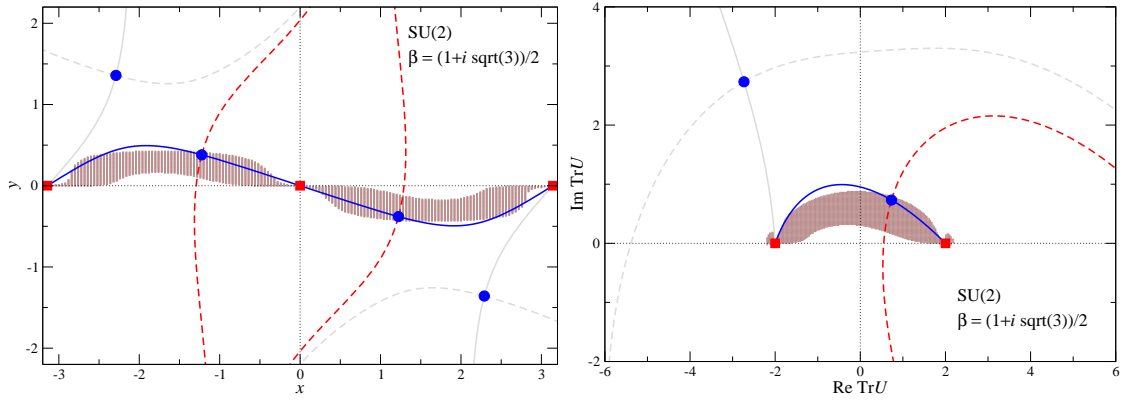

\begin{center}
\epsfig{figure=plot-SU2-thimble-xy-CL.eps,width=0.48\textwidth}
\epsfig{figure=plot-SU2-thimble-action-CL.eps,width=0.48\textwidth}
\end{center}
\caption{
Comparison between the complex Langevin histograms and the thimbles, in the $xy$ plane (left) and the $\Tr U$ plane (right).
}
 \label{fig:SU2-th-CL}
\end{figure}

We now compare the thimbles with the results from a complex Langevin simulation.
 Fig.\  \ref{fig:SU2-CL}  shows the histograms collected during a simulation in the $xy$ plane (left) and the $\Tr U$ plane (right). We note that the distributions are localised; this model and its complex Langevin dynamics was discussed in detail in Ref.\ \cite{Aarts:2013uxa}. A comparison between the thimbles and the histograms is shown in Fig.\  \ref{fig:SU2-th-CL}. As in  previous cases, we note that the Langevin distribution is close to the contributing thimble, but  two rather than one-dimensional, to be able to evade the residual phase problem. From a comparison with the exact results, we conclude that correct results are obtained both with Langevin dynamics and with the single contributing thimble, provided the residual phase is included.

\subsection{Gauge dynamics with cooling}

The logarithm in the gauge-fixed action and the resulting singular drift arise from the reduced Haar measure after gauge fixing. It is therefore interesting to apply Langevin dynamics without gauge fixing, using the matrix update
\be
U(t+\eps) = R(t) U(t), \quad\quad\quad R = \exp\left [i\sigma_a\left(\eps K_a+\sqrt{\eps}\eta_a\right)\right].
\ee
Here $t$ is the (discretised) Langevin time, $K_a=-D_a S$ is the drift and $\sigma_a$ are the Pauli matrices \cite{Aarts:2008rr}. It is by now well established that a straightforward application of this approach can lead to wrongly converging Langevin dynamics, due to an uncontrolled exploration of SL(2,$\mathbb{C}$), the complexification of SU(2). Hence it needs to be combined with gauge cooling \cite{Seiler:2012wz}, so that the combined update reads
\be
U(t+\eps) = \Omega(t) R(t) U(t) \Omega^{-1}(t), 
\ee
where $\Omega(t)$ is a $U(t)$ dependent SL(2,$\mathbb{C}$) gauge transformation, which minimises the distance from SU(2)
\cite{Seiler:2012wz,Aarts:2013uxa}.

\begin{figure}[t]
\begin{center}
\epsfig{figure=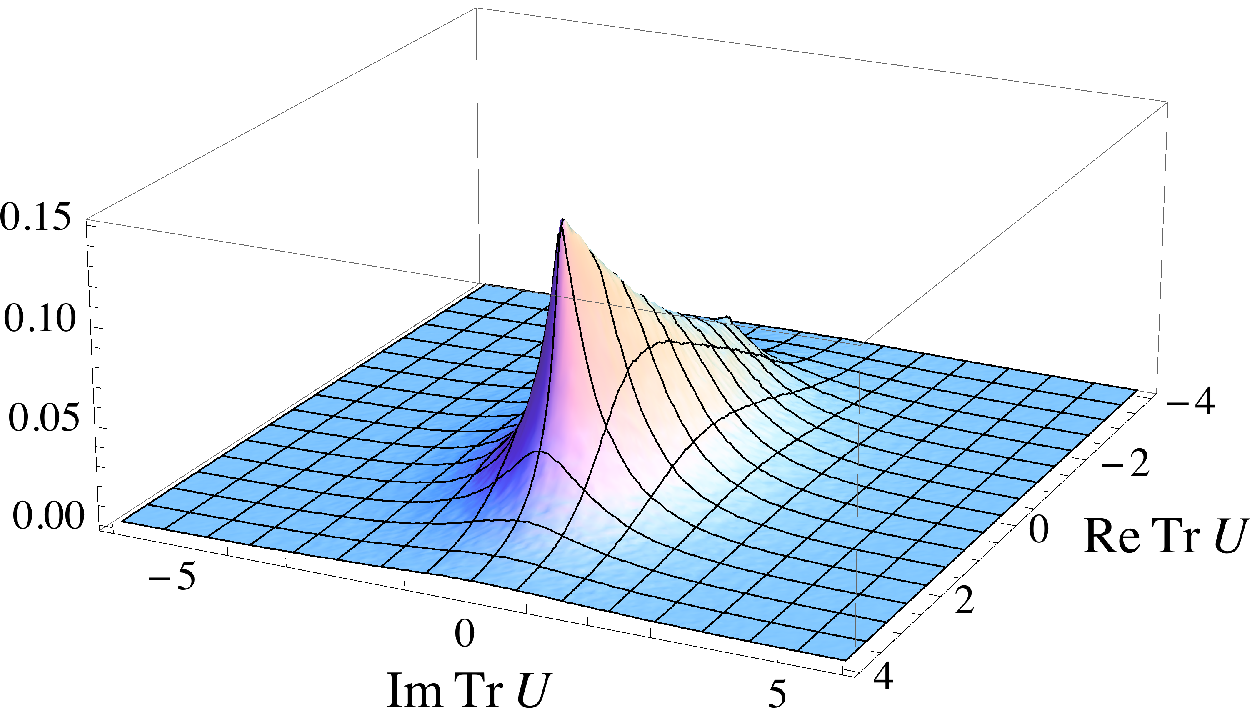,width=0.48\textwidth}
\epsfig{figure=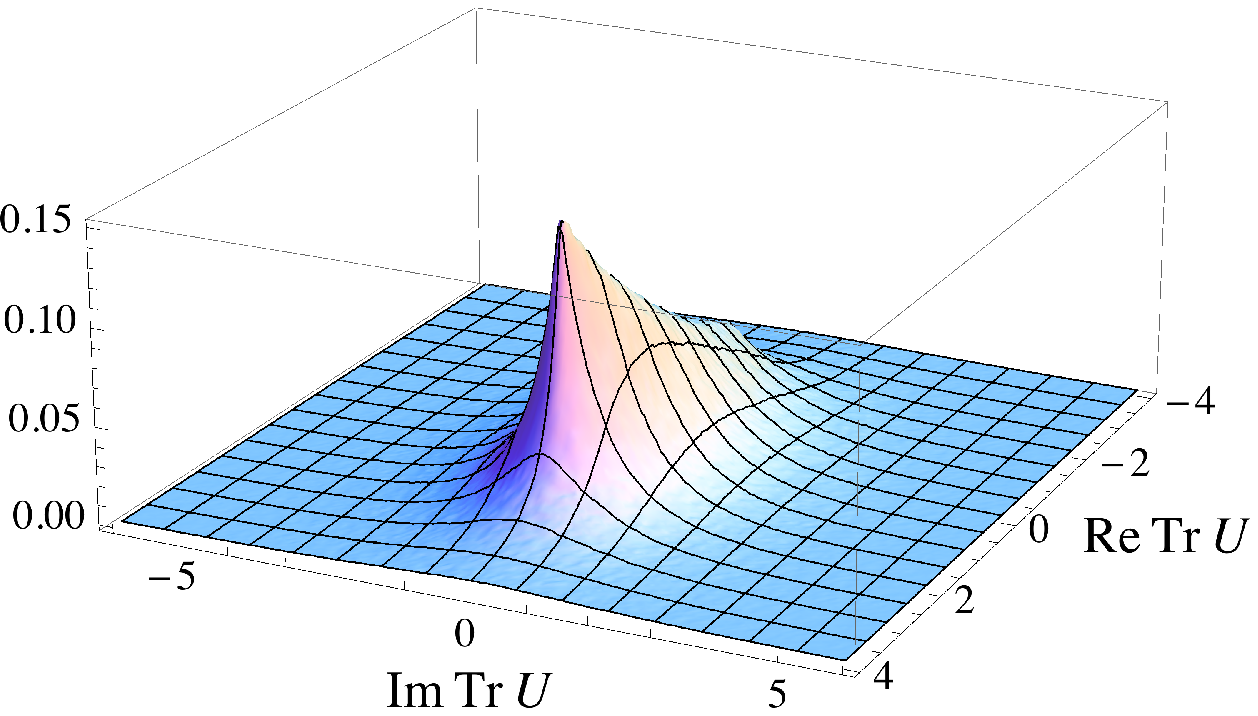,width=0.48\textwidth}
\epsfig{figure=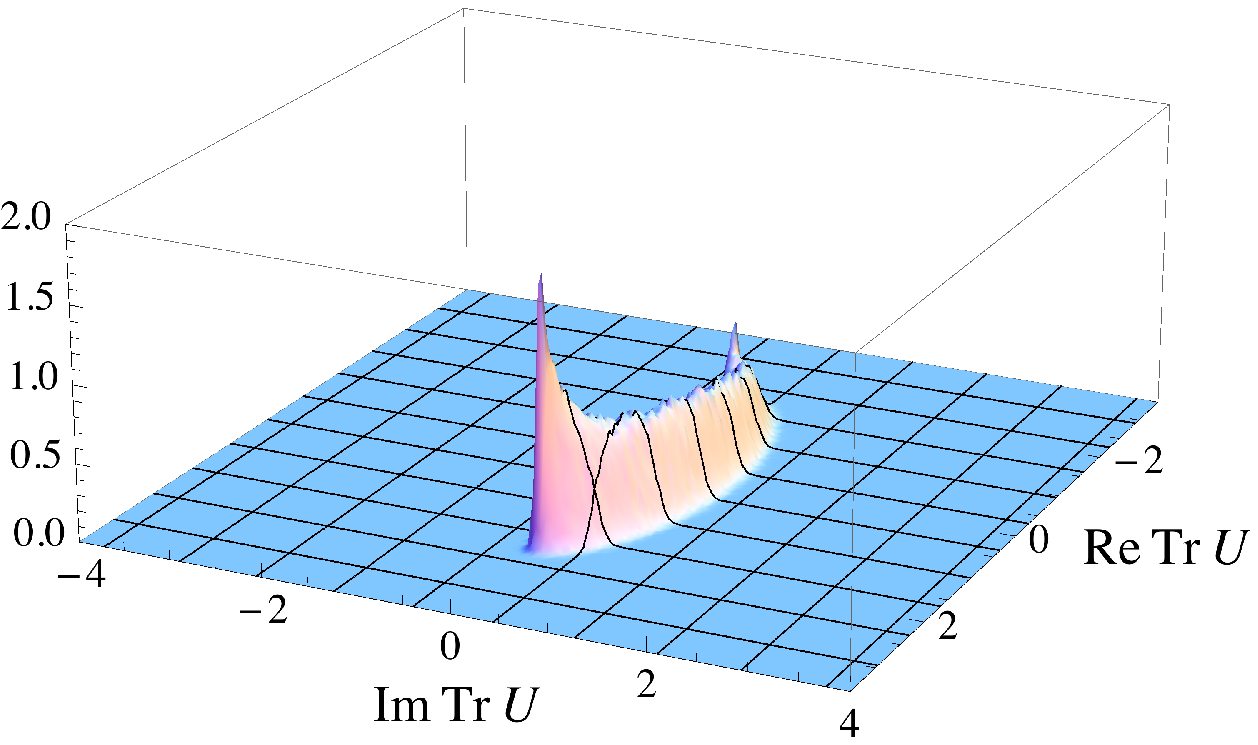,width=0.48\textwidth}
\epsfig{figure=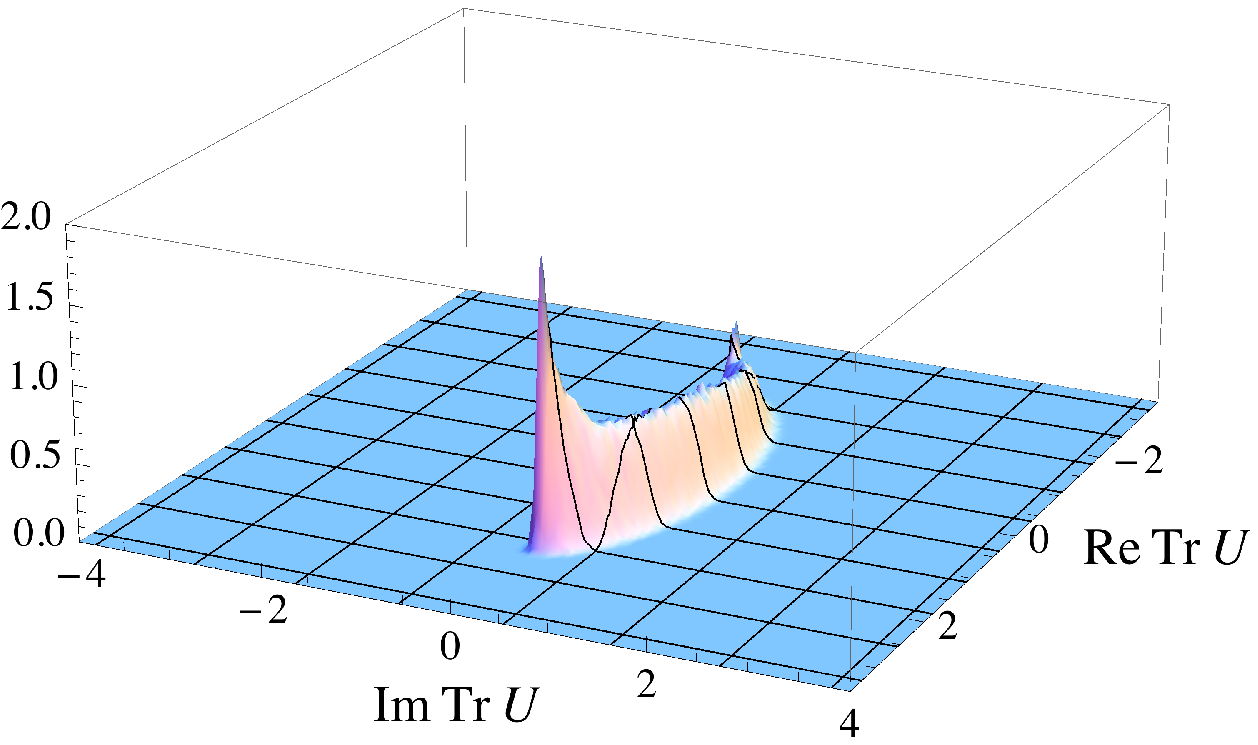,width=0.48\textwidth}
\end{center}
\caption{
Histograms collected during a complex Langevin simulation in the matrix formulation of the SU(2) one-link model at $\beta=(1+i\sqrt{3})/2$, with gauge cooling, using 0, 1, 2, 4 (from top left to bottom right) gauge cooling steps.}
 \label{fig:SU2-CL-cooling}
\end{figure}

In a simulation, the number of cooling steps applied after each Langevin update can be varied to achieve stability. 
In Fig.\ \ref{fig:SU2-CL-cooling} we show the results of a Langevin simulation using 0, 1, 2 or 4 gauge cooling steps between each Langevin update. Without cooling the distribution is broad and wrong results are obtained. One gauge cooling step is insufficient to cure this. Perhaps surprisingly, applying just 2 cooling steps results in a drastically different distribution, which is now well localised in the $\Tr U$ plane. According to the justification of the approach \cite{arXiv:0912.3360,arXiv:1101.3270}, correct results should now be obtained, which is indeed the case. Moreover, subsequent cooling does not modify the distribution any further (the bottom left plot is for 4 cooling steps), which is a sign of stability and convergence. Finally, we have compared the distributions in the $\Tr U$ plane obtained in the gauge fixed formulation and the matrix formulation, and found agreement.
We conclude therefore that the dynamics is under complete control and that the singular drift in the first approach and the need for gauge cooling in the second approach do not affect Langevin dynamics in a detrimental manner.
Moreover, the position of the distribution is related to the position of the thimble. The distribution of the weight is, however, again quite distinct.

\subsection{Degenerate fixed points}
\label{sec43}

When $\beta^2=-1$ (and hence $\beta$ is purely imaginary\footnote{The choice of imaginary $\beta$ is motivated by dynamics in real (Minkowskian) time, see e.g.\ Refs.\ \cite{Berges:2005yt,Berges:2006xc,Berges:2007nr} for complex Langevin studies.}),
  the fixed point at  $z_d$, with $\cos z_d=-1/\beta=\pm i$, is degenerate. The standard reasoning to justify the Lefschetz approach and construct the thimbles by numerical integration is then not well defined. Here we give a brief analysis.

At the fixed point the action is real, with $S(z_d) = 1-\ln 2$.
We take $\beta=i$ and write
\be
\half\Tr U = \cos z = u+i v.
\ee
Equating the imaginary part of the action,
\be
\im S = -u-\phi, \quad\quad\quad \tan\phi = \frac{-2uv}{1-u^2+v^2},
\ee
to 0, then yields the thimbles, and we find
\be
v_\pm(u) = \frac{1}{\tan u}\left(u\pm\sqrt{u^2-(1-u^2)\tan^2u}\right),
\ee 
where  $0<u<1$ for $v_-(u)$ and $-1<u<0$ for $v_+(u)$. These two branches make up the stable thimble. 
The unstable thimble is given by $u=0$, for which the imaginary part of the action vanishes as well.

\begin{figure}[t]
\begin{center}
\epsfig{figure=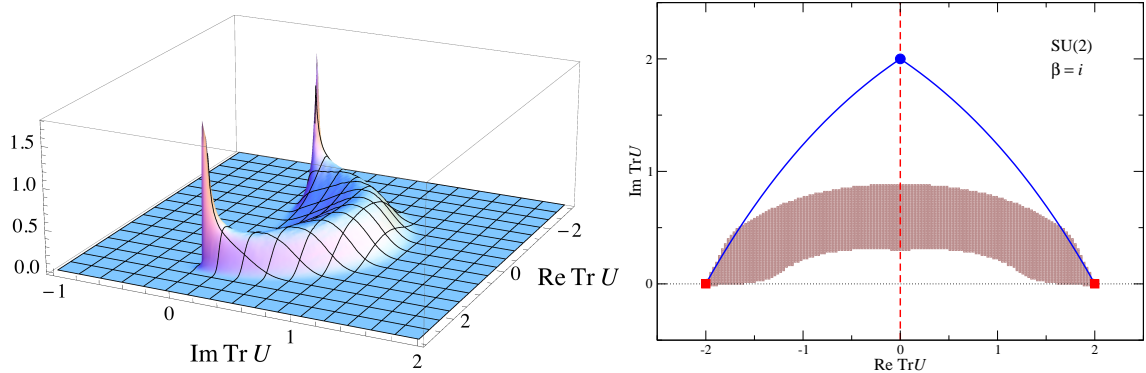,width=0.53\textwidth}
\epsfig{figure=plot-SU2-thimble-beta-i.eps,width=0.45\textwidth}
\end{center}
\caption{
SU(2) one-link model at $\beta=i$. Histogram collected during a complex Langevin simulation in the  $\Tr U$ plane (left) and a comparison with the thimbles associated with the degenerate fixed point at $\Tr U=2i$ (right).
}
 \label{fig:SU2-th2-beta-i}
\end{figure}

The thimbles are shown in Fig.\  \ref{fig:SU2-th2-beta-i} (right), using the same colour coding as above. They are a deformation of the thimbles for general complex $\beta$, shown earlier, satisfying reflection symmetry in the $\re\Tr U=0$ axis. We have verified that integrating along the thimble, with the inclusion of the residual phase, gives the correct answer.
The Langevin histogram is shown in Fig.\ \ref{fig:SU2-th2-beta-i} (left). A similar histogram was obtained earlier in Ref.\ \cite{Berges:2007nr}. 
A comparison between the histogram and the thimbles is finally given in Fig.\ \ref{fig:SU2-th2-beta-i} (right). 
For this case we note that the distribution does not overlap substantially with the thimble and that again the distribution of the weight for both approaches is quite different.

\section{Summary and outlook}
\label{sec:conc}

We have studied and contrasted the distributions sampled by complex Langevin dynamics and the Lefschetz thimbles in a number of models with a complex action, focussing on examples with a determinant in the Boltzmann weight. Due to the determinant the drift is non-holomorphic and has singular points, which leads to new features in both the Langevin and the thimble approach. In particular, thimbles may end at the singularities and hence do not necessarily extend to infinity. Moreover, it naturally leads to a situation where more thimbles need to be taken into account.

In general, there is substantial overlap between the regions explored in the complexified configuration space using Langevin and thimble dynamics. However, a  closer look reveals important differences. The weight on a (one-dimensional) thimble is peaked around the fixed point and drops quickly to zero away from the fixed point. On the other hand, the two-dimensional probability distribution sampled in complex Langevin dynamics is often peaked away from fixed points. A further important  difference is due to the presence of repulsive fixed points in Langevin dynamics. In the thimble case  these correspond to saddle points and the thimbles passing through those may contribute. However, in Langevin dynamics the region around repulsive fixed points is avoided. It therefore follows that the thimble and the Langevin distributions will be manifestly different around repulsive fixed points.
Finally, the justification of Langevin dynamics for holomorphic actions depends on the decay of the distribution in the imaginary direction.\footnote{It is 
sometimes suggested that failure of Langevin dynamics is due to the existence of several probability distributions which may or may not be sampled simultaneously  and that each of those corresponds to a distinctive solution of the Schwinger-Dyson equations (SDEs), subject to particular boundary conditions \cite{Salcedo:1993tj,Pehlevan:2007eq,Guralnik:2009pk}. However, in the models and field theories we considered, this does not seem to occur. In fact, we usually find a unique stationary distribution, robust against variation of initial conditions, stepsize and other ingredients of the numerical algorithm (the exception is for non-abelian gauge theories, in the case that gauge cooling is required to control the process). 
In our experience the origin of failure of complex Langevin for holomorphic actions lies in a slowly decaying probability distribution $P(x,y)$ and consequently the breakdown of partial integration required for the formal justification \cite{arXiv:0912.3360}. The slow decay impacts the SDEs as well, since higher-order moments are no longer well-defined. In Ref.\ \cite{Aarts:2013uza} this was demonstrated in an explicit example, both numerically and analytically.}
 Hence contributing thimbles going to $z\to x\pm i\infty$ are potentially dangerous, since they open up a way to prevent the distribution from being localised. In particular, they correspond to classical runaway trajectories.
However, here it should be noted that the weight on those thimbles will be exponentially small far away from saddle points, making the relation between a possible breakdown of Langevin dynamics and the relevant region in thimble dynamics  less immediate.

The thimble approach has to face a residual and a global sign problem. 
For the simple models we considered, the correct inclusion of the residual phase on the thimbles is mandatory. In field theory, on the other hand, it appears that the residual sign problem can in fact be quite mild. One possibility is that the weight on the thimble drops to zero away from the saddle point in an exponential fashion, with a rate set by the volume. 
In that case the thimble is nearly flat in the contributing region, which makes the residual phase problem milder than perhaps expected from the simple models. It would be worthwhile to investigate this further.
A global sign problem appears when more than one thimble contributes. 
This possibility is highlighted in the Stokes phenomenon, in which a transition from, say, one to two contributing thimbles takes place, as the external parameters are changed. It would be interesting to see how important it is to capture this transition accurately in field theory, where there is typically less analytical control.

Finally, the most important outstanding question for complex Langevin dynamics is the role of the determinant and the loss of holomorphicity, which may or may not lead to a breakdown of the approach in practice. Finding a precise answer to this is currently under investigation.

%***************************************************************

 \vspace*{0.5cm}
 \noindent
 {\bf Acknowledgments} \\
\noindent
We thank Ion-Olimpiu Stamatescu for discussion.
ES and DS thank Swansea University for hospitality during the course of this work. GA is supported by STFC, the Royal Society, the Wolfson Foundation and the Leverhulme Trust. ES is supported by Deutsche Forschungsgemeinschaft.

%***************************************************************

%\appendix
%\label{sec:app}

\end{document}